\documentclass[times,twocolumn]{aastex631}
\usepackage{graphicx, amsmath}
\usepackage{booktabs}
\usepackage{threeparttable}

\newcommand{\ha}{\text{H}\alpha}

\newcommand{\hd}{\text{H}\delta}
\newcommand{\dn}{\text{D}_{n}4000}
\newcommand{\haew}{\text{EW}(\text{H}\alpha)}
\newcommand{\hdew}{\text{EW}(\text{H}\delta_A)}
\newcommand{\dhaew}{\Delta \text{EW}(\text{H}\alpha)}
\newcommand{\dhdew}{\Delta \text{EW}(\text{H}\delta_A)}

\newcommand{\pdis}{d_{\text{p}}}

\newcommand{\logm}{\log M_\star}

\newcommand{\logsfr}{\log \text{SFR}}
\newcommand{\ashape}{A_{\text{shape}}}

\newcommand{\kpc}{\ \text{kpc}}

\newcommand{\kms}{\ \text{km}\ \text{s}^{-1}}

\shorttitle{Star Formation Evolution in Galaxy Pairs}
\shortauthors{Geng et al.}

\graphicspath{{./}{Figure/}}

\begin{document}

\title{Star Formation Evolution in Galaxy Pairs: Constraints from Morphological Disturbances and Recent Star Formation Histories}

\correspondingauthor{Shuai Feng}
\email{sfeng@hebtu.edu.cn}

\author[0009-0009-4187-2095]{Shaoze Geng}
\affiliation{College of Physics, Hebei Normal University, 20 South Erhuan Road, Shijiazhuang 050024, China}
\affiliation{University of Chinese Academy of Sciences (UCAS), Beijing 100049, China}

\author[0000-0002-9767-9237]{Shuai Feng}
\affiliation{College of Physics, Hebei Normal University, 20 South Erhuan Road, Shijiazhuang 050024, China}
\affiliation{Shijiazhuang Key Laboratory of Astronomy and Space Science, Shijiazhuang 050024, China}
\affiliation{Hebei Key Laboratory of Photophysics Research and Application, Shijiazhuang 050024, China}

\author{Shuhao Chen}
\affiliation{College of Physics, Hebei Normal University, 20 South Erhuan Road, Shijiazhuang 050024, China}

\author{Yaotian Duan}
\affiliation{College of Physics, Hebei Normal University, 20 South Erhuan Road, Shijiazhuang 050024, China}
\affiliation{College of Physics, Dalian University of Technology, 2 Linggong Road, Dalian 116024, China}

\author[0000-0002-3073-5871]{Shiyin Shen}
\affiliation{Shanghai Astronomical Observatory, Chinese Academy of Sciences, 80 Nandan Road, Shanghai 200030, China}
\affiliation{Key Lab for Astrophysics, Shanghai 200234, China}

\author[0009-0005-3397-2038]{Cailu Shi}
\affiliation{College of Physics, Hebei Normal University, 20 South Erhuan Road, Shijiazhuang 050024, China}

\author[0000-0003-1454-2268]{Linlin Li}
\affiliation{College of Physics, Hebei Normal University, 20 South Erhuan Road, Shijiazhuang 050024, China}
\affiliation{Shijiazhuang Key Laboratory of Astronomy and Space Science, Shijiazhuang 050024, China}

\author{Yijin Niu}
\affiliation{College of Physics, Hebei Normal University, 20 South Erhuan Road, Shijiazhuang 050024, China}

\author{Bingxi Huo}
\affiliation{College of Physics, Hebei Normal University, 20 South Erhuan Road, Shijiazhuang 050024, China}

\author[0000-0003-1359-9908]{Wenyuan Cui}
\affiliation{College of Physics, Hebei Normal University, 20 South Erhuan Road, Shijiazhuang 050024, China}
\affiliation{Shijiazhuang Key Laboratory of Astronomy and Space Science, Shijiazhuang 050024, China}

\author[0000-0003-1828-5318]{Guozhen Hu}
\affiliation{College of Physics, Hebei Normal University, 20 South Erhuan Road, Shijiazhuang 050024, China}
\affiliation{Shijiazhuang Key Laboratory of Astronomy and Space Science, Shijiazhuang 050024, China}

\begin{abstract}
Galaxy interactions can enhance star formation, but how star formation evolves during galaxy interactions remains poorly constrained by observations. We combine pair separation, morphological disturbance, and recent star formation history to study this evolution. We measure morphological disturbance with the shape asymmetry parameter ($\ashape$) using deep images from the DESI Legacy Surveys, and use Galaxy Zoo DESI classifications as an independent test. We infer recent star formation histories by comparing SFR and $\haew$, which trace current star formation, with $\hdew$ and $\dn$, which are sensitive to stellar populations formed over longer timescales. At small projected separations, strongly disturbed galaxies show the strongest enhancements in SFR and $\haew$, indicating that close encounters trigger strong star formation. At intermediate separations ($\pdis\sim100\kpc$), their current star formation is only moderately enhanced, but their $\hdew$ enhancement is the strongest. This indicates a larger contribution from intermediate-age stars formed during stronger star formation in the past $\sim0.1{-}1$ Gyr. The TNG100 analysis shows that this pattern results from rapid changes in SFR around close passage. SFR rises sharply during the encounter and declines as the galaxies move apart, while tidal disturbances remain visible. As short-lived massive stars disappear and intermediate-age A-type stars begin to dominate the spectrum, the $\hdew$ enhancement peaks later than the SFR enhancement. Our results provide important observational evidence for the evolution of SFR along the merger sequence.
\end{abstract}


\section{Introduction} \label{sec:intro}

Galaxy mergers play a fundamental role in galaxy evolution, driving transformations in galaxy morphology and star formation activity \citep[e.g.,][]{Barnes1996, Hopkins2008}. Because galaxy pairs represent systems in the early stages of interaction before final coalescence, they provide an important opportunity to investigate how galaxy interactions affect galaxies during the merging process. Compared with isolated galaxies of similar stellar mass, galaxy pairs exhibit enhanced star formation activity \citep{Barton2000, Ellison2008, Li2008} and more disturbed morphologies \citep{HernandezToledo2005, Casteels2013, Li2026}. Many of these effects become progressively stronger toward smaller projected pair separations \citep{Patton2013, Patton2016, Feng2024}, suggesting that tidal interactions play a major role in driving these phenomena. 

Numerical simulations suggest that many of the unusual properties observed in galaxy pairs originate from tidal disturbances during close encounters \citep[e.g.,][]{DiMatteo2007, Cox2008, Lawrence2025}. During pericentric passages, strong tidal torques can perturb stellar structures and drive gas inflows toward galaxy centers, triggering enhanced star formation and producing disturbed morphologies \citep{Mihos1996, Torrey2012, Bustamante2018}. Simulations further predict that galaxy mergers proceed through multiple orbital passages before final coalescence. After a close encounter, the galaxies may move apart toward larger separations before reapproaching each other for subsequent interactions \citep[e.g.,][]{Springel2005, Lotz2010a, Patton2024}. As a result, the strength of tidal interaction evolves throughout the merger process, leading to corresponding stage-dependent variations in galaxy properties such as morphology and star formation activity \citep{Lotz2008, Montuori2010, Sparre2022}. Understanding how star formation evolves across different merger stages is therefore essential for revealing the physical impact of galaxy interactions.

Observationally, several studies have attempted to reconstruct how star formation evolves throughout the merger process by comparing galaxy properties at different merging stages. Motivated by numerical simulations, these studies generally infer merger stage using a combination of projected pair separation and tidal disturbance indicators, such as morphological asymmetry \citep{Haan2011, Pan2019, Jin2021} or kinematic disturbance \citep{Feng2020, Yu2022, Yu2024}. Within this simulation-motivated framework, previous studies have generally interpreted galaxy pairs with small projected separations and strong tidal disturbances as systems near pericentric passage, while pairs at larger separations but still exhibiting disturbance features are often associated with post-passage stages. In contrast, relatively undisturbed pairs are usually interpreted as systems before the first close encounter. By comparing the star formation properties of systems classified into different merger stages, previous studies have attempted to reconstruct the evolutionary sequence of interaction-induced star formation \citep{Shangguan2019, Calderon-Castillo2024, Pearson2025}.

However, this approach remains subject to significant uncertainties. Projected pair separation is affected by projection effects and cannot directly trace the true three-dimensional distance between galaxies \citep{Perez2006, Contreras-Santos2022, Patton2024}. More importantly, the physical separation between merging galaxies does not evolve monotonically throughout the interaction process. As a result, systems with similar projected separations may correspond to very different evolutionary stages. At the same time, the interpretation of tidal disturbance features is itself strongly model-dependent. Numerical simulations predict that morphological disturbances evolve throughout mergers, but the detailed evolution depends sensitively on orbital configuration, gas fraction, mass ratio, intrinsic galaxy structure, viewing angle, and dynamical relaxation processes \citep{Lotz2010a, Lotz2010b, McElroy2022}. Consequently, the evolutionary behavior of star formation throughout the merger process remains poorly constrained observationally.

To better constrain how star formation evolves during galaxy mergers, independent observational tracers of the merger stage are required. Unlike projected separation and tidal disturbance, whose interpretations depend strongly on orbital geometry and numerical simulations, spectral diagnostics sensitive to different star formation timescales provide direct constraints on the recent star formation history of galaxies. For example, $\ha$ emission traces ongoing star formation on short timescales \citep{FloresVelazquez2021, Tacchella2022}, while $\hd$ absorption and $\dn$ are sensitive to stellar populations formed over longer periods \citep{Worthey1997, Kauffmann2003}. Because these indicators respond differently to variations in star formation activity, their combined behavior provides insight into the recent evolution of star formation in galaxies \citep[e.g.,][]{Chen2019, Wang2020, Iyer2024}. Such multi-timescale spectral diagnostics therefore offer an observationally independent approach for reconstructing the evolutionary sequence of interaction-induced star formation.

In this work, we investigate the connection between morphological disturbance, projected pair separation, and recent star formation history in a large sample of star-forming galaxy pairs. By combining these observables, we aim to better constrain how star formation evolves throughout the merger process. We further compare the observational results with numerical simulations to place the observed trends within a physical evolutionary framework. This paper is organized as follows. In Section~\ref{sec:data}, we describe the galaxy pair sample and observational data. Section~\ref{sec:disturb} presents the measurements of morphological disturbance and the corresponding statistical properties. In Section~\ref{sec:rSFH}, we investigate the connection between morphology, projected pair separation, and recent star formation history. Section~\ref{sec:dis} compares the observational results with TNG100 simulations to explore the evolution of star formation during galaxy interactions. Finally, Section~\ref{sec:sum} summarizes our main conclusions. Throughout this work, we adopt a standard $\Lambda$CDM cosmology with $H_0 = 70~\mathrm{km~s^{-1}~Mpc^{-1}}$, $\Omega_{\rm m}=0.3$, and $\Omega_\Lambda=0.7$.

\section{Data} \label{sec:data}

\subsection{Galaxy Pair Sample} \label{sec:pair_sample}

The galaxy pairs analyzed in this study are selected from the main galaxy sample of SDSS DR7 \citep{Strauss2002, Blanton2005, SDSSDR7}, which contains approximately 750,000 galaxies with $r<17.77$. Because of spectroscopic incompleteness (e.g., fiber collisions and limitations in fiber assignment), about $60,000$ galaxies lack spectroscopic redshifts. To improve the spectroscopic completeness, we supplement the missing redshifts using additional spectroscopic surveys \citep[see details in][]{Shen2016, Feng2019}, including SDSS DR16 \citep{SDSSDR16}, LAMOST DR10 \citep{Zhao2012, Luo2015}, and GAMA DR3 \citep{Driver2011, Baldry2018}. The combined catalog reaches an overall spectroscopic completeness of $96.3\%$. At angular separations smaller than $55''$, where the fiber-collision effect is most significant, the spectroscopic completeness increases from $\sim30\%$ in the original SDSS DR7 catalog \citep{Liu2011, Patton2016} to $\sim65\%$ after including the additional spectroscopic surveys. At $z=0.1$, $55''$ corresponds to a projected separation of approximately $100$ kpc, demonstrating that the supplemented catalog substantially improves the spectroscopic completeness on the physical scales relevant for close-pair identification.

Galaxy pairs are identified from the parent sample using the following criteria: (1) both galaxies satisfy $0.02<z<0.12$; (2) the line-of-sight velocity difference satisfies $|\Delta v| \leq 500\kms$; (3) the projected separation satisfies $10\kpc \leq \pdis \leq 200\kpc$; and (4) each pair member has exactly one companion satisfying the above criteria, ensuring that the selected systems are isolated galaxy pairs rather than galaxy groups.

Several additional constraints are further applied to construct the final sample. First, we restrict the analysis to spiral--spiral pairs in order to minimize the influence of hot circumgalactic gas associated with elliptical galaxies on star formation activity \citep{Feng2024, Shi2026}. Spiral galaxies are identified using the S\'ersic index, requiring $n<2.5$, where the S\'ersic indices are taken from the catalog of \citet{Simard2011}. Second, we focus on major-merger systems by requiring the stellar mass ratio to satisfy $1/3 \leq M_A/M_B \leq 3$, and both galaxies to have stellar masses within $9.5 < \log(M_*/M_\odot) < 11.0$. Stellar masses are primarily adopted from the MPA--JHU catalog \citep{Kauffmann2003}. For galaxies without MPA--JHU measurements, stellar masses are estimated using the $g-r$ color relation of \citet{Bell2003}. Third, because this work focuses on the recent star formation histories of interacting galaxies, we further restrict the sample to star-forming galaxies. Star-forming galaxies are identified using the BPT diagram based on the emission-line ratios [\ion{O}{3}]$/\mathrm{H\beta}$ and [\ion{N}{2}]$/\mathrm{H\alpha}$, adopting the demarcation of \citet{Kauffmann2003BPT, Kewley2006}. Finally, all candidate pairs are visually inspected using optical images in order to remove spurious systems and obvious image artifacts. The final sample contains approximately $5090$ star-forming pair-member galaxies, which are used in all subsequent analyses.

\begin{figure*}
  \centering
  \includegraphics[width=\textwidth]{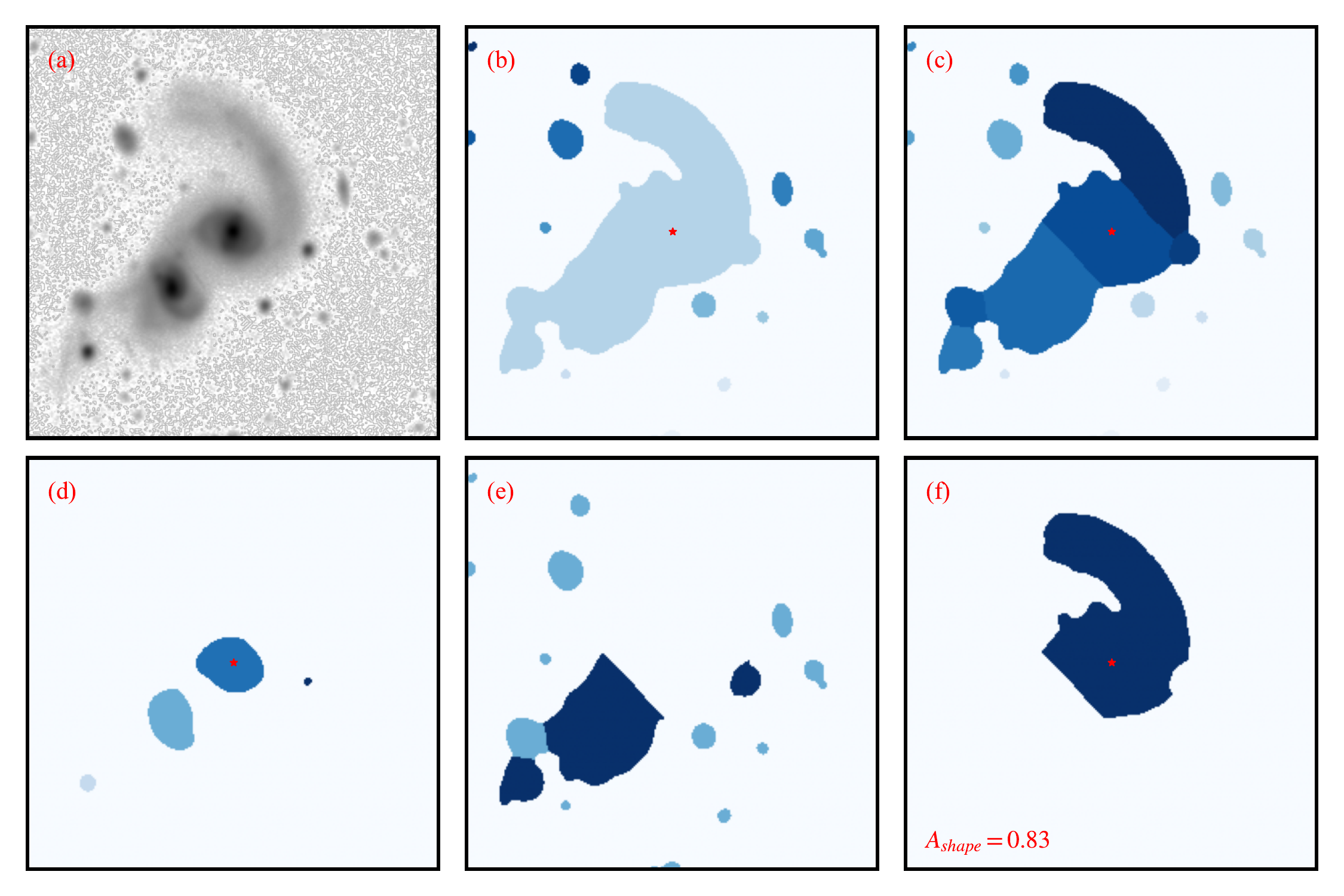}
  \caption{Illustration of the $\ashape$ measurement procedure. Panel (a) shows the Legacy Survey $r$-band image. Panels (b), (c), and (d) present the source detection results obtained from the cold-mode segmentation, cold-mode deblending, and hot-mode deblending procedures, respectively. Panel (e) shows the mask constructed from pixels associated with contaminating sources surrounding the target galaxy. Panel (f) presents the final pixels assigned to the target galaxy and used for the $\ashape$ measurement. The measured $\ashape$ value is labeled in the lower-left corner of panel (f).}
  \label{fig:flow_chart}
\end{figure*}

\subsection{Photometric Images} \label{sec:image}

We use photometric images from the tenth public data release (DR10) of the DESI Legacy Imaging Surveys\footnote{\url{https://www.legacysurvey.org/dr10/}} for our morphological analysis. The survey combines imaging data from three projects conducted with different telescopes: the Beijing--Arizona Sky Survey \citep[BASS;][]{Zou2017}, the Mayall $z$-band Legacy Survey (MzLS), and the DECam Legacy Survey (DECaLS). As of DR10, the combined survey covers more than 20,000 square degrees across both the Northern and Southern Galactic Caps, providing $g$, $r$, and $z$-band imaging with $5\sigma$ depths of 24.7, 23.9, and 23.0 mag, respectively \citep{Dey2019}. Compared to SDSS imaging, the significantly greater depth of the Legacy Surveys enables the detection of faint low-surface-brightness structures associated with galaxy interactions, such as tidal tails, asymmetric outer features, and diffuse tidal debris.

For each galaxy, we extract an $r$-band image cutout centered on the target galaxy. When both members of a galaxy pair fall within the same cutout, separate cutouts are generated for each galaxy to ensure that the target galaxy remains centered in the image. The morphological analysis is then performed only on the central target galaxy, while surrounding sources, including the companion galaxy and nearby background objects, are treated as contaminating sources during the source-detection and masking procedures.

Figure~\ref{fig:flow_chart}(a) presents an example of a galaxy pair in the Legacy Surveys. The target galaxy is centered in the frame and exhibits a prominent tidal tail extending toward the upper right, indicative of strong tidal disturbance. The companion galaxy, located toward the lower left, also shows faint asymmetric structures. Several nearby background galaxies are additionally visible in the field. This example demonstrates the capability of the Legacy Surveys imaging to reveal faint tidal features in interacting galaxies.

\section{Morphological Disturbance Measurements} \label{sec:disturb}

Morphological disturbances produced by galaxy interactions can be identified using several approaches, including visual classification, machine-learning-based methods, and non-parametric morphological indicators. Visual classification is the most direct and intuitive way to identify disturbed systems, particularly for recognizing complex tidal features such as tails, bridges, and shells \citep[e.g.,][]{Haan2011, Pan2019, Calderon-Castillo2024a}. However, traditional visual inspection is relatively inefficient and can be affected by differences among individual researchers, limiting its application to very large samples.

Machine-learning methods enable the identification of interacting and merging galaxies in large samples. These methods generally follow two training strategies. The first uses visually classified galaxies as training data and extends visual morphology classifications to much larger samples \citep[e.g.,][]{Walmsley2023, Ye2025}. The second uses simulated galaxy images labeled according to their merger histories \citep[e.g.,][]{Bickley2021, Ferreira2024, Margalef-Bentabol2024}.

Non-parametric methods quantify galaxy structure using image statistics. Common indicators include the Concentration--Asymmetry--Smoothness (CAS) system \citep{Conselice2003}, Gini--$M_{20}$ statistics \citep{Lotz2004}, outer asymmetry \citep{Wen2014, Ren2024}, and shape asymmetry \citep{Pawlik2016}. These methods are efficient and reproducible, making them suitable for quantifying morphological disturbances in large galaxy samples.

In this work, we primarily use a non-parametric morphology indicator to quantify morphological disturbance, and use machine-learning-based morphology classifications trained on visual labels as an independent comparison.

\begin{figure*}
    \centering
    \includegraphics[width=\textwidth]{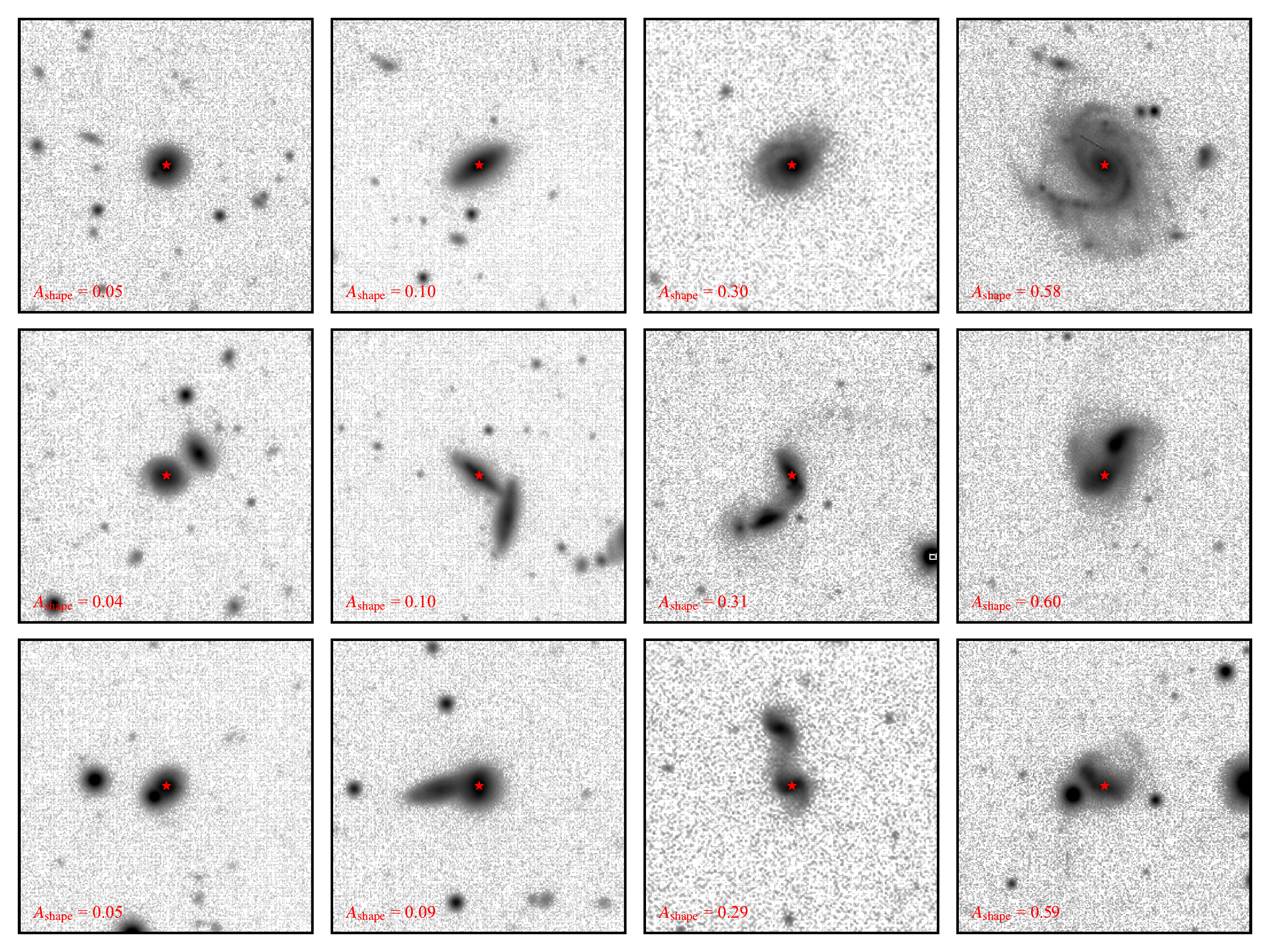}
    \caption{Legacy Survey $r$-band cutout images of galaxy pairs. The target galaxy is marked with a red asterisk in each panel, and its corresponding value of shape asymmetry is shown in the lower-left corner.}
    \label{fig:AshapeCase}
\end{figure*}

\subsection{Shape Asymmetry} \label{sec:ashape}

The shape asymmetry parameter \citep[$\ashape$;][]{Pawlik2016} is designed to quantify the asymmetry of the spatial distribution of luminous pixels in a galaxy image. Unlike traditional light-weighted asymmetry measurements, $\ashape$ is computed from a binary segmentation map in which all detected pixels are assigned equal weight regardless of their surface brightness. Consequently, $\ashape$ is particularly sensitive to faint low-surface-brightness structures in the outer regions of galaxies, such as tidal tails, shells, and disturbed outskirts, while being less affected by bright internal structures, including bars and spiral arms.

However, measuring $\ashape$ in close galaxy pairs is challenging. Overlapping light from companion galaxies can contaminate the segmentation boundary and artificially enhance the measured asymmetry, while standard source-detection methods often either blend neighboring galaxies together or fragment faint tidal features into multiple disconnected structures. To overcome these issues, we adopt a cold+hot detection strategy implemented with the \texttt{photutils} package, following previous studies \citep[e.g.,][]{Galametz2013, Sazonova2021, Zhao2022}. This approach is designed to simultaneously preserve faint tidal structures and improve the separation between close companion galaxies.

We first perform a cold-mode detection on the $r$-band image shown in Figure~\ref{fig:flow_chart}(a). Before detection, the image is smoothed with a Gaussian kernel of $\sigma=3$ pixels. Sources are then identified down to a surface-brightness threshold of $25.5~\mathrm{mag\,arcsec^{-2}}$, allowing faint tidal structures and diffuse outer features to be recovered. The resulting cold-mode segmentation map is shown in Figure~\ref{fig:flow_chart}(b). We then further deblend the detected structures by identifying distinct local brightness peaks. A relatively conservative deblending criterion is adopted, in which a secondary peak is separated only if it contains at least $10\%$ of the total flux of the blended structure. The deblended cold-mode result is shown in Figure~\ref{fig:flow_chart}(c). At this stage, both companion galaxies and tidal structures may be separated into individual components.

To distinguish companion galaxies from diffuse tidal features, we additionally perform a hot-mode detection using a higher surface-brightness threshold of $23~\mathrm{mag\,arcsec^{-2}}$ together with a much more aggressive deblending criterion, allowing secondary peaks with flux contrasts down to $10^{-6}$ to be separated. Because the hot-mode detection is primarily sensitive to compact high-surface-brightness structures, it efficiently identifies the central regions of neighboring galaxies while remaining relatively insensitive to faint tidal features. The resulting hot-mode deblending map is shown in Figure~\ref{fig:flow_chart}(d). By comparing the cold- and hot-mode detections, structures identified in the cold-mode segmentation can be classified as either companion galaxies or tidal features.

The final segmentation map is then constructed based on this classification. Cold-mode structures identified as tidal features are retained as part of the target galaxy, while structures associated with companion galaxies or nearby contaminating sources are excluded. Specifically, a cold-mode structure is assigned to the target galaxy only if it contains pixels within twice the Petrosian radius of the target, allowing nearby tidal structures to remain associated with the target galaxy. Structures outside this radius are treated as belonging to companion galaxies or other nearby sources.

Based on the final classification, we construct a shape map in which pixels associated with the target galaxy are assigned a value of 1, while all other pixels are set to 0. We also construct a mask map containing pixels associated with companion galaxies and contaminating sources. The mask map, shown in Figure~\ref{fig:flow_chart}(e), is used to exclude overlapping regions during the $\ashape$ measurement, thereby minimizing artificial asymmetry introduced by companion light contamination. The final pixels assigned to the target galaxy and used for the $\ashape$ measurement are shown in Figure~\ref{fig:flow_chart}(f).

The shape asymmetry is then defined as
\begin{equation}
    A_{\text{shape}} =
    \frac{
    \sum_{i,j} M_{ij}\, M_{ij}^{\mathrm{rot}}
    \left| I_{ij}-I_{ij}^{\mathrm{rot}} \right|
    }{
    \sum_{i,j} M_{ij}\, M_{ij}^{\mathrm{rot}} I_{ij}
    },
\end{equation}
where $I_{ij}$ is the value of the shape map at pixel $(i,j)$ and $I_{ij}^{\mathrm{rot}}$ is the corresponding value after a $180^\circ$ rotation. The mask map is denoted by $M_{ij}$, where unmasked pixels have $M_{ij}=1$ and masked pixels have $M_{ij}=0$. The term $M_{ij}\, M_{ij}^{\mathrm{rot}}$ ensures that pixels affected by companion galaxies or other contaminating sources are excluded consistently from both the original and rotated shape maps.

The shape asymmetry parameter ranges from $0.001$ to $1.58$. Figure~\ref{fig:AshapeCase} presents $12$ example galaxies and galaxy pairs together with their measured $\ashape$ values. In each panel, the target galaxy is marked by the red asterisk. The examples show that $\ashape$ is strongly correlated with the prominence of tidal disturbance features in the target galaxy. Systems with clear tidal tails, asymmetric outer structures, or disturbed outskirts generally exhibit larger $\ashape$ values, while morphologically regular galaxies show much lower asymmetry. In addition, several very close galaxy pairs still exhibit relatively small $\ashape$ values when the target galaxy itself shows little evidence of tidal disturbance. These examples demonstrate that $\ashape$ primarily traces the intrinsic tidal disturbance of the target galaxy rather than simply the presence of a nearby companion.

\subsection{Galaxy Zoo DESI Visual Morphology} \label{sec:gzd}

In addition to the non-parametric morphology measurements, we use machine-learning-based morphology classifications trained on visual classifications. For this purpose, we adopt the Galaxy Zoo DESI (GZ DESI) catalog \citep{Walmsley2023}. GZ DESI extends the earlier Galaxy Zoo DECaLS project \citep{Walmsley2022} to the full footprint of the DESI Legacy Imaging Surveys and provides detailed morphology measurements for approximately $8.7$ million galaxies. These classifications are generated using deep-learning models trained on volunteer classifications from both the Galaxy Zoo DECaLS project and newly collected Galaxy Zoo classifications for the DESI Legacy Imaging Surveys.

In Galaxy Zoo, volunteers classify galaxies by answering a series of morphology questions, each requiring the selection of one response from several possible options. The fraction of volunteers selecting each response, referred to as the vote fraction, provides a quantitative description of the corresponding morphological feature. GZ DESI predicts these vote fractions for each galaxy and morphology question. For a given question, the predicted vote fractions of all possible responses sum to unity, although several or even all responses can have non-zero values for an individual galaxy.

In this work, we focus specifically on the interaction-related morphology question, which includes four possible responses: \textsc{merger}, \textsc{merger\_none}, \textsc{minor\_disturb}, and \textsc{major\_disturb}. The \textsc{merger} response corresponds primarily to galaxies with a very close or overlapping companion in the image, while the \textsc{major\_disturb} and \textsc{minor\_disturb} responses describe galaxies exhibiting strong and weak morphological disturbances, respectively. The \textsc{merger\_none} response corresponds to galaxies without obvious close companions or visible disturbance features.

To facilitate comparison with $\ashape$, we combine these four responses into three broader categories: (i) \textit{none}, corresponding to \textsc{merger\_none}; (ii) \textit{merger}, corresponding to \textsc{merger}; and (iii) \textit{disturb}, defined as the combination of \textsc{minor\_disturb} and \textsc{major\_disturb}. For each galaxy, the fractions of the original four classes are combined accordingly, and we assign the galaxy to the category with the highest combined fraction for our analysis.

It should be noted that the GZ DESI disturbance classifications do not distinguish between morphological disturbances produced by ongoing interactions and those associated with post-merger remnants. Therefore, the \textit{disturb} category may include some post-merger remnants. For our galaxy pair sample, the requirement of a spectroscopic companion excludes isolated post-merger remnants without current companions, and the disturbances are more likely to be associated with ongoing or recent interactions involving the pair members. For the isolated control sample, a small number of post-merger remnants may still be present. However, observationally identified post-mergers constitute only a small fraction of the low-redshift galaxy population \citep[e.g.,][]{Ellison2013}. Their contribution is therefore unlikely to significantly affect the median properties used to define the control baseline.

\begin{figure*}
    \centering
    \includegraphics[width=\linewidth]{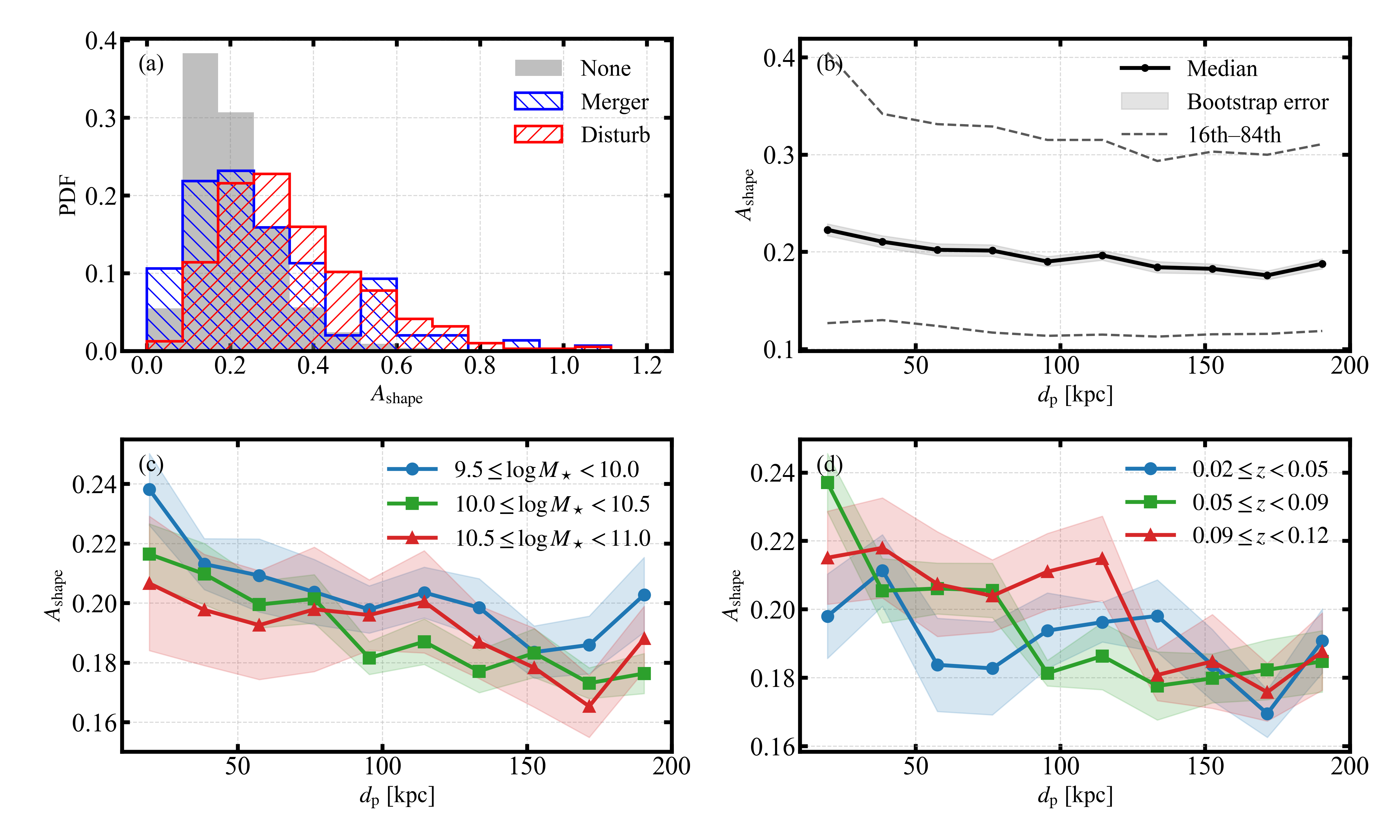}
    \caption{
    (a) Normalized distributions of $\ashape$ for galaxies classified as none, merger, and disturb, shown in gray, blue, and red, respectively.
    (b) Median $\ashape$ as a function of projected separation ($\pdis$). The gray shaded region shows the $1\sigma$ bootstrap uncertainty of the median from $1000$ resamplings, while the dashed lines mark the 16th and 84th percentiles of the $\ashape$ distribution.
    (c) Median $\ashape$ as a function of $\pdis$ in different stellar-mass bins; shaded regions show the $1\sigma$ bootstrap uncertainties.
    (d) Same as panel (c), but for different redshift bins.
    }
    \label{fig:AshapeProperty}
\end{figure*}

Figure~\ref{fig:AshapeProperty}(a) shows the distributions of $\ashape$ for galaxies in the three visual classes. Galaxies in the \textit{none} class generally exhibit low $\ashape$ values, with more than half of the sample having $\ashape < 0.2$. In contrast, galaxies in the \textit{disturb} class show systematically higher asymmetry, with the majority having $\ashape > 0.2$, indicating that large $\ashape$ values are statistically associated with visually identified tidal disturbances. 

However, the distributions also exhibit substantial overlap. Part of this overlap is related to limitations in the $\ashape$ measurement itself. For example, faint satellite galaxies or residual contamination may occasionally be identified as asymmetric structures, leading to artificially enhanced $\ashape$ values in otherwise undisturbed systems. Conversely, in strongly disturbed galaxies, some diffuse tidal features may be fragmented or incompletely recovered during the segmentation process, resulting in relatively low measured asymmetry. Despite these effects, the overall statistical trend remains clear: galaxies with stronger visually identified disturbances tend to exhibit systematically larger $\ashape$ values.

The \textit{merger} class has a broad $\ashape$ distribution because the Galaxy Zoo \textit{merger} response primarily identifies a close or overlapping companion rather than the strength of tidal disturbance. Since \textit{merger}, \textit{major disturbance}, and \textit{minor disturbance} are alternative responses to the same interaction-related morphology question, the GZ DESI vote fractions do not provide an independent measure of tidal disturbance for galaxies assigned to the \textit{merger} class. This class may therefore include strongly disturbed interacting systems, weakly disturbed close pairs, and a small fraction of projected pairs that may not be physically interacting. In this work, the GZ DESI classifications are used as an independent qualitative comparison, while $\ashape$ provides a continuous quantitative measure that allows morphological disturbance to be examined separately from projected pair separation.

\subsection{Dependence of Shape Asymmetry} \label{sec:AshapeDependence}

Figure~\ref{fig:AshapeProperty}(b) shows the distribution of $\ashape$ as a function of projected separation $\pdis$. The black solid line represents the median $\ashape$ in each bin, with the gray shaded region showing the $1\sigma$ bootstrap uncertainty estimated from $1000$ resamplings, while the dashed lines indicate the $16$th--$84$th percentile range. The $\ashape$ distribution is broad at all separations, indicating substantial diversity in galaxy morphology within each $\pdis$ bin. Nevertheless, a clear statistical trend is present, with the median $\ashape$ increasing systematically toward smaller projected separations. This behavior is consistent with the expectation that tidal interactions become progressively stronger at small separations and further supports the interpretation of $\ashape$ as a tracer of interaction-induced morphological disturbance. The upper envelope of the distribution also increases significantly toward small $\pdis$, indicating that highly asymmetric systems become substantially more common in close galaxy pairs. At the same time, a considerable fraction of galaxies still exhibit low $\ashape$ values even at very small projected separations, suggesting that close projected pairs include a non-negligible population of systems that are not yet strongly interacting.

Figures~\ref{fig:AshapeProperty}(c) and (d) further examine the dependence of the $\ashape$--$\pdis$ relation on stellar mass and redshift, respectively. In Figure~\ref{fig:AshapeProperty}(c), the shaded regions represent the uncertainties on the median estimated via bootstrap resampling. All stellar-mass subsamples exhibit similar behavior, with $\ashape$ increasing toward smaller projected separations. No significant dependence on stellar mass is observed over the range $9.5 < \logm < 11$. Similarly, Figure~\ref{fig:AshapeProperty}(d) shows that the $\ashape$--$\pdis$ relation remains broadly consistent across the redshift range $0.02 < z < 0.12$, with no strong systematic differences between redshift subsamples. This result suggests that observational effects such as spatial resolution do not significantly bias the $\ashape$ measurement within our sample.

Taken together, these results indicate that $\ashape$ provides a robust statistical tracer of interaction-induced tidal disturbance, with only weak dependence on stellar mass and redshift over the parameter range explored in this work. This weak dependence further suggests that a single $\ashape$ threshold can be used to statistically characterize the strength of tidal disturbance across our sample.

\section{Morphological Disturbance and Recent Star Formation History} \label{sec:rSFH}

\subsection{Star Formation Diagnostics and Timescales}

To characterize the recent star formation history (rSFH) of galaxies, we employ several spectral diagnostics that respond to star formation activity over different characteristic timescales. The H$\alpha$ equivalent width ($\haew$) traces ongoing star formation on short timescales of $\lesssim 10$ Myr because it is powered by ionizing photons from massive young stars \citep{FloresVelazquez2021, Tacchella2022}. The equivalent width of H$\delta$ absorption ($\hdew$) is sensitive to stellar populations formed within the past $\sim0.1{-}1$ Gyr, and becomes enhanced when A-type stars contribute significantly to the optical spectrum \citep{Worthey1997, Wang2020}. The $4000\,\text{\AA}$ break strength ($\dn$) reflects the luminosity-weighted stellar age and is reduced by the presence of young stellar populations, with its response persisting over longer timescales than H$\alpha$ \citep{Kauffmann2003}.

Because these diagnostics respond differently to changes in star formation activity, their relative behavior provides information about the temporal evolution of recent star formation \citep[e.g.,][]{Wang2020, Iyer2024}. For example, during the rising phase of star formation, galaxies are expected to exhibit enhanced $\haew$ together with reduced $\dn$, while $\hdew$ may not yet show strong enhancement. After the star formation activity begins to decline, $\haew$ can decrease rapidly, whereas $\hdew$ may remain elevated because of the increasing contribution from intermediate-age A-type stars. The $\dn$ index also evolves during this process, but its variation is typically smoother and persists over longer timescales because it traces the cumulative contribution of young and intermediate-age stellar populations to the optical continuum.

Galaxy interactions are expected to produce strongly time-dependent star formation histories through tidal perturbations, gas inflows, and subsequent gas consumption or relaxation \citep{Cox2008, Torrey2012}. As a result, galaxies at different merging stages may exhibit distinct combinations of these diagnostics. By jointly analyzing projected separation, morphological disturbance, and multi-timescale star formation indicators, these spectral signatures can therefore help us interpret how star formation may evolve across different merger stages.

\subsection{Recent Star Formation History as a Function of Shape Asymmetry}\label{sec:AshapeSFH}

\begin{figure*}
  \centering
  \includegraphics[width=\textwidth]{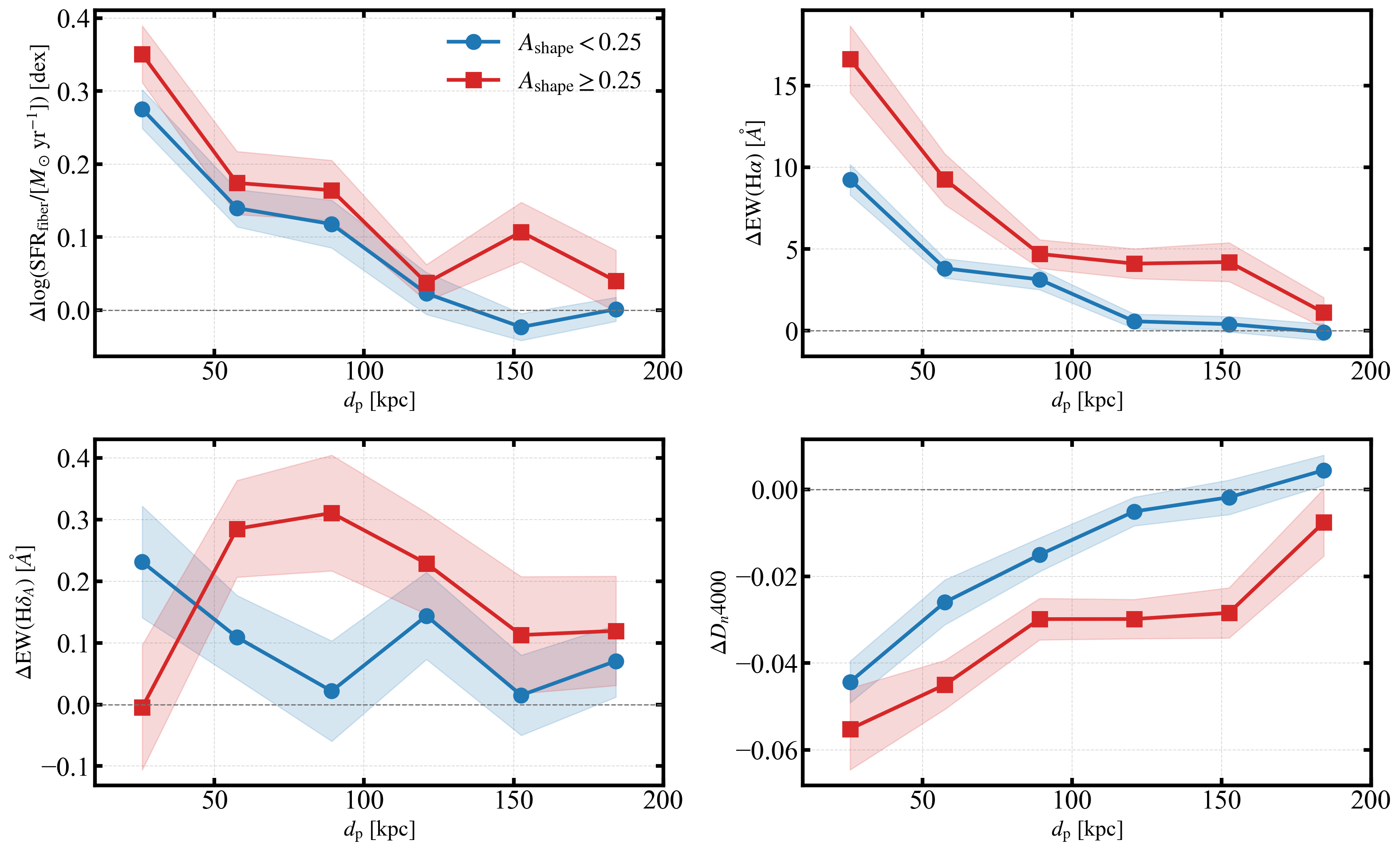}
  \caption{Star formation properties of galaxy pairs as a function of projected separation ($\pdis$), shown as offsets relative to a mass- and redshift-matched control sample. Panels show (from top-left to bottom-right) fiber SFR, $\dhaew$, $\dhdew$, and $\Delta D_n4000$. Red squares and blue circles represent galaxies with $\ashape \geq 0.25$ and $\ashape < 0.25$, respectively. Symbols show median values in each bin, with errors estimated via bootstrap resampling.}
  \label{fig:AshapeSFR}
\end{figure*}

To characterize recent star formation history over different timescales, we use four spectral diagnostics measured within the SDSS $3^{''}$ fiber aperture, including the fiber star formation rate (SFR), the equivalent widths of $\ha$ and $\hd$ ($\haew$ and $\hdew$), and the $4000\,\AA$ break strength ($\dn$). For our sample at $0.02<z<0.12$, the $3^{''}$ fiber aperture corresponds to physical diameters of approximately $1$--$7$ kpc, and therefore primarily probes the central regions of galaxies rather than their integrated properties. To quantify the interaction-induced variations in recent star formation history, we construct a control sample of isolated galaxies matched in stellar mass ($|\Delta \logm| < 0.1$) and redshift ($|\Delta z| < 0.01$) for each galaxy in the pair sample. The median values of the corresponding diagnostics in the control sample are adopted as baselines. We then define the offsets relative to these controls as $\Delta \logsfr$, $\Delta \haew$, $\Delta \hdew$, and $\Delta \dn$, which quantify the enhancement or suppression of central recent star formation associated with galaxy interactions.

Figure~\ref{fig:AshapeSFR} presents the dependence of these offsets on projected separation ($\pdis$) for galaxy pairs with different $\ashape$. Galaxies with $\ashape \geq 0.25$ are shown as red squares, while those with $\ashape < 0.25$ are shown as blue circles. The solid lines represent the median values in each projected-separation bin, and the shaded regions indicate the $1\sigma$ uncertainties on the medians estimated through bootstrap resampling. 

The top panels show the trends of $\Delta \logsfr$ and $\Delta \haew$, both of which trace current star formation activity and therefore exhibit broadly similar behavior. Overall, both diagnostics remain close to the control level at large projected separations and increase rapidly once $\pdis \lesssim 100\kpc$. This trend is broadly consistent with previous observational studies of galaxy pairs, which have found enhanced star formation toward smaller projected separations, with detectable enhancement in some samples extending to separations beyond $100\kpc$ \citep[e.g.,][]{Ellison2008, Li2008, Patton2013}. When the pair sample is further divided according to morphological disturbance, a clear difference emerges. Galaxies with high $\ashape$ show stronger enhancement in both $\Delta \logsfr$ and $\Delta \haew$, and the enhancement extends to larger projected separations than in the low-$\ashape$ population. In contrast, galaxies with low $\ashape$ show significant enhancement mainly at small separations. These results demonstrate that projected separation alone does not fully capture the interaction-induced star formation response, and that the degree of morphological disturbance provides additional information on the evolutionary state of interacting galaxies.

The bottom-left panel shows the behavior of $\Delta \hdew$, which traces star formation over intermediate timescales of $\sim0.1{-}1$ Gyr. Galaxies with low $\ashape$ show little variation in $\Delta \hdew$ over most of the projected separation range, with values remaining close to zero and therefore comparable to the control sample. A modest enhancement is seen only at the smallest separations ($\pdis \lesssim 30\kpc$). In contrast, galaxies with high $\ashape$ show a different behavior. At large separations ($\pdis \gtrsim 150\kpc$), $\Delta \hdew$ remains close to zero and shows no significant difference from either the control sample or the low-$\ashape$ population. However, at intermediate separations ($30 \lesssim \pdis \lesssim 150\kpc$), high-$\ashape$ galaxies exhibit enhanced $\Delta \hdew$ values relative to the control sample, with the strongest enhancement occurring at intermediate separations around $50$--$100\kpc$. In this range, $\Delta \hdew$ is also significantly higher than that of the low-$\ashape$ population. At the smallest separations ($\pdis \lesssim 30\kpc$), $\Delta \hdew$ in high-$\ashape$ galaxies decreases back to a level comparable to the control sample and becomes lower than that of the low-$\ashape$ population.

This behavior differs markedly from the trends observed in $\logsfr$ and $\haew$, where the strongest enhancement occurs at the smallest projected separations. The strong $\Delta \hdew$ enhancement at intermediate separations in the high-$\ashape$ population suggests that these systems experienced strong star formation activity over the past $\sim0.1{-}1$ Gyr. In contrast, the weaker $\Delta \hdew$ enhancement at the smallest separations, despite strong enhancement in $\logsfr$ and $\haew$, suggests that the current star formation enhancement in these systems is relatively recent and has not yet produced a strong H$\delta$ absorption signature. Meanwhile, the increase of $\Delta \hdew$ at small projected separations in the low-$\ashape$ population indicates that these galaxies also experienced enhanced star formation over the past $\sim0.1{-}1$ Gyr, although with substantially weaker tidal disturbances. These differences show that current and past star formation do not vary in the same way with projected separation and morphological disturbance.

The bottom-right panel shows the behavior of $\Delta \dn$, which traces changes in the luminosity-weighted stellar age. Galaxies with low $\ashape$ show a relatively simple trend: $\Delta \dn$ remains close to zero at large projected separations and decreases steadily toward smaller $\pdis$, indicating progressively younger stellar populations. In contrast, galaxies with high $\ashape$ display a more complex dependence on separation. At relatively large separations, $\Delta \dn$ shows evidence for reduced values in high-$\ashape$ galaxies, particularly in the $150$--$175\kpc$ bin, suggesting that some strongly disturbed systems may already host younger stellar populations even at relatively large projected separations. At intermediate separations ($75 \lesssim \pdis \lesssim 175\kpc$), $\Delta \dn$ remains relatively flat, while at smaller separations ($\pdis \lesssim 75\kpc$), it decreases again toward smaller $\pdis$. This behavior suggests that the stellar populations of some high-$\ashape$ galaxies may already have been affected before they reach the smallest projected separations, suggesting that enhanced star formation had already occurred in some of these systems.

Taken together, these results show that galaxies with different combinations of projected separation and morphological disturbance have different recent star formation histories. Strongly disturbed galaxies at the smallest separations show the strongest current star formation, while their relatively weak $\hdew$ enhancement suggests that this activity is recent. At intermediate separations, strongly disturbed galaxies show weaker current star formation but stronger $\hdew$ enhancement and reduced $\dn$, indicating stronger star formation in the past $\sim0.1{-}1$ Gyr and a larger contribution from younger stellar populations. Galaxies with low $\ashape$ generally show weaker variations, although those at the smallest separations exhibit evidence of enhanced current and recent star formation.

\subsection{Recent Star Formation History as a Function of Galaxy Zoo Morphology}

\begin{figure*}
  \centering
  \includegraphics[width=\textwidth]{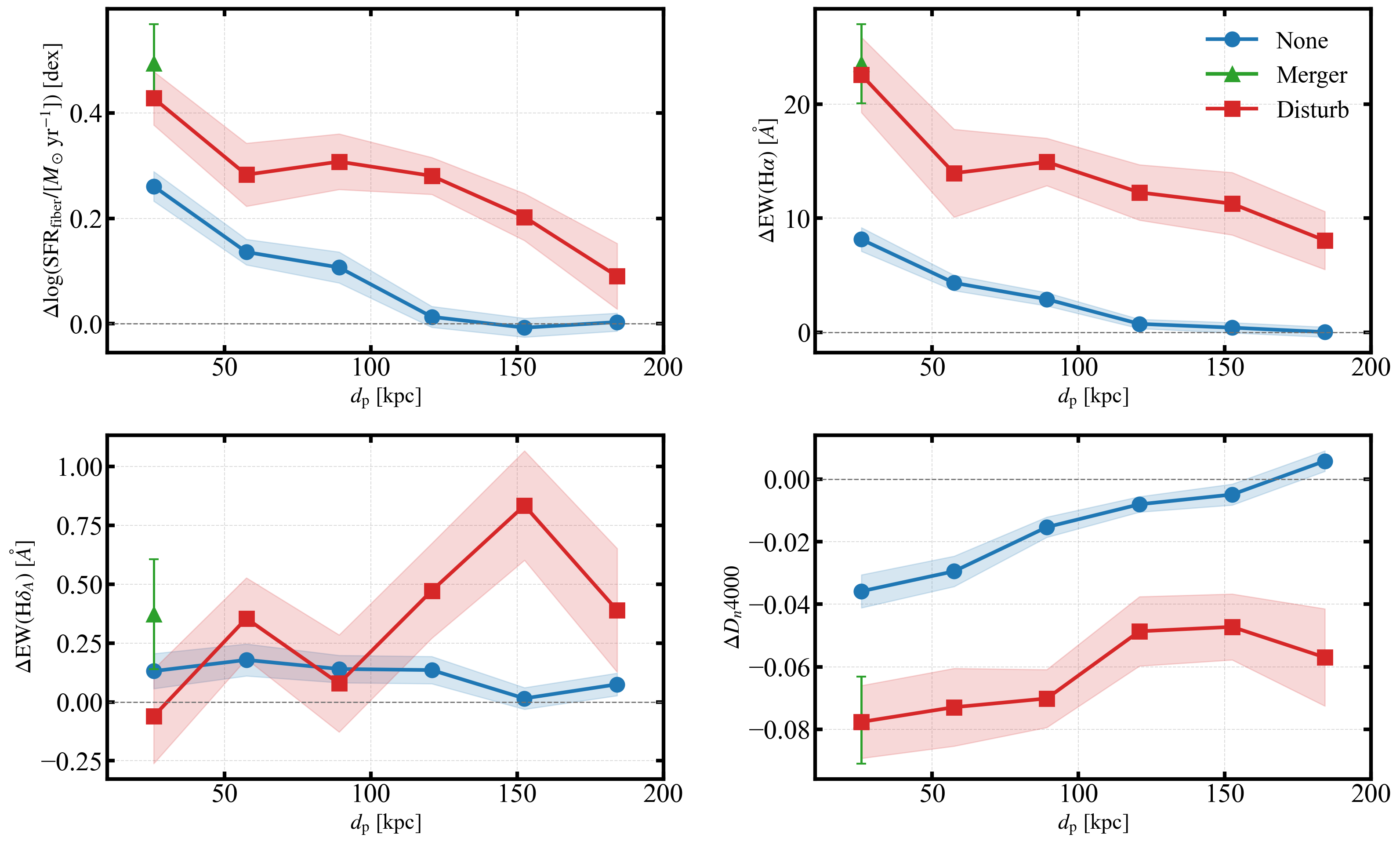}
  \caption{Same as Figure~\ref{fig:AshapeSFR}, but with galaxies classified according to Galaxy Zoo morphology. Blue circles, green triangles, and red squares represent \textit{none}, \textit{merger}, and \textit{disturb}, respectively.}
  \label{fig:GzSFH}
\end{figure*}

To test whether the trends found with $\ashape$ depend on the adopted morphology indicator, we repeat the analysis using the GZ DESI classifications introduced in Section~\ref{sec:gzd}. Galaxies are divided into three morphology classes based on the GZ DESI predictions: \textit{none}, \textit{merger}, and \textit{disturb}. Figure~\ref{fig:GzSFH} shows the star formation diagnostics as a function of projected separation for these three classes. Blue circles, green triangles, and red squares represent the \textit{none}, \textit{merger}, and \textit{disturb} classes, respectively.

The overall trends are consistent with those obtained using $\ashape$. Galaxies in the \textit{disturb} class show stronger enhancements in $\logsfr$ and $\haew$ than galaxies in the \textit{none} class, particularly at small projected separations. The trends in $\hdew$ and $\dn$ show similar agreement. The \textit{disturb} class exhibits a stronger reduction in $\dn$ than the other classes, while a significant enhancement in $\hdew$ is found only in the \textit{disturb} class and mainly at relatively large projected separations. These results indicate that disturbed galaxies at larger separations experienced stronger star formation in the recent past.

The detailed dependence on projected separation differs between the two morphology indicators. In the $\ashape$ analysis, the strongest $\hdew$ enhancement in high-$\ashape$ galaxies mainly occur at $\pdis\sim75\kpc$. In the GZ DESI analysis, the \textit{disturb} class shows enhanced $\hdew$ over a broader separation range, extending to $\pdis\sim150\kpc$. This difference mainly reflects the different strengths of tidal disturbance selected by the two classifications. The GZ DESI \textit{disturb} class preferentially selects galaxies with prominent tidal features, whereas the adopted threshold of $\ashape\geq0.25$ also includes galaxies with relatively weak disturbances. When a higher $\ashape$ threshold is adopted, the resulting trends become consistent with those of the GZ DESI \textit{disturb} class.

The \textit{merger} class generally shows behavior intermediate between the \textit{none} and \textit{disturb} classes. As discussed in Section~\ref{sec:gzd}, the GZ DESI \textit{merger} response identifies a close or overlapping companion but does not independently quantify its tidal disturbance. This class therefore includes close pairs with a wide range of disturbance strengths, as well as a small fraction of projected pairs, which likely accounts for its intermediate behavior. Overall, the agreement between the GZ DESI and $\ashape$ analyses supports the connection between morphological disturbance and recent star formation history.

\subsection{Past Star Formation Activity at Fixed Current Star Formation} \label{sec:rSFH_fixHA}

\begin{figure*}
  \centering
  \includegraphics[width=\textwidth]{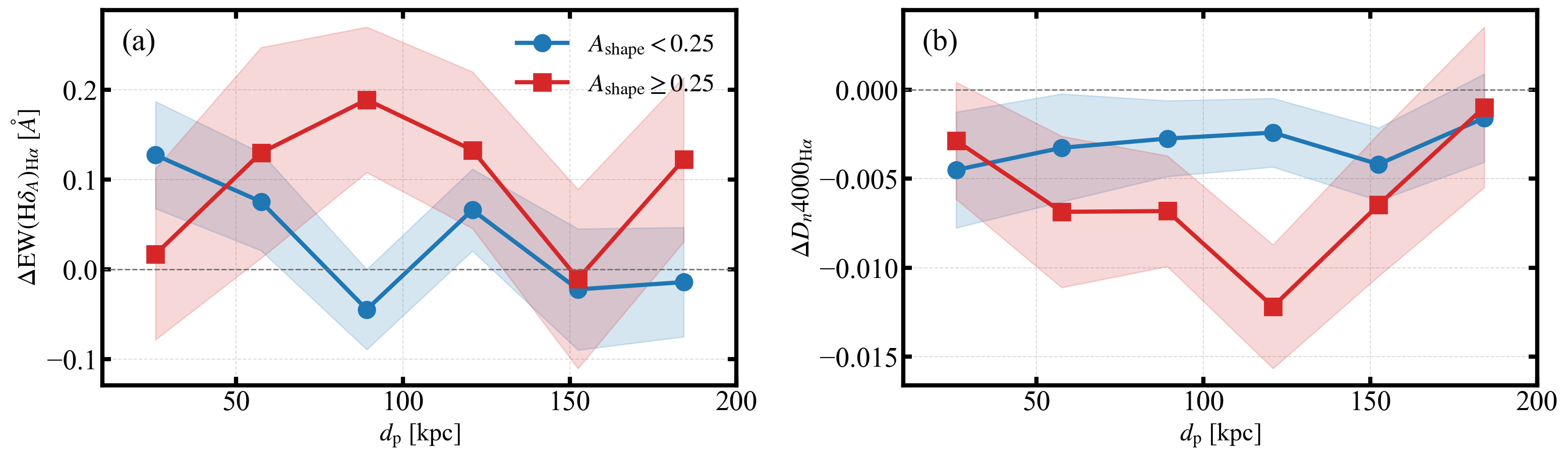}
  \caption{Offsets of $\dhdew_{\rm H\alpha}$ and $\Delta D_n4000_{\rm H\alpha}$ as a function of projected separation ($\pdis$), shown relative to a control sample additionally matched in H$\alpha$ equivalent width. Blue circles and red squares represent galaxies with $\ashape < 0.25$ and $\ashape \geq 0.25$, respectively. }
  \label{fig:dHDdD4K}
\end{figure*}

To further isolate the impact of interactions on recent star formation history, we construct an alternative control sample with an additional constraint on the current star formation activity. In addition to matching in stellar mass ($\Delta \logm < 0.1$) and redshift ($\Delta z < 0.01$), we also require a close match in $\haew$ ($\Delta \log \haew < 0.05$), ensuring that galaxies in the control sample have similar levels of ongoing star formation as those in the pair sample. In this way, the control sample provides a reference for isolated galaxies with comparable present-day star formation activity.

For clarity, we denote the offsets relative to this control sample as $\dhdew_{\ha}$ and $\Delta \dn_{\ha}$. Since the control galaxies already have similar current star formation activity, any residual differences in these quantities indicate differences in past star formation activity between pair and isolated galaxies at fixed current star formation. Figure~\ref{fig:dHDdD4K} presents the dependence of $\dhdew_{\ha}$ and $\Delta \dn_{\ha}$ on projected separation.

The left panel shows the behavior of $\dhdew_{\ha}$. Galaxies with low $\ashape$ remain broadly consistent with the isolated control sample over most projected separations, indicating similar star formation activity over the past $\sim0.1{-}1$ Gyr at fixed current star formation. A mild enhancement is seen at the smallest separation ($\pdis<30\kpc$), suggesting that these close pairs may have experienced somewhat stronger star formation in the recent past than isolated galaxies. In contrast, galaxies with high $\ashape$ show significant enhancement in $\dhdew_{\ha}$ at intermediate separations, indicating stronger star formation over the past $\sim0.1{-}1$ Gyr. The largest-separation bin also shows a trend toward enhanced $\dhdew_{\ha}$. Combined with the results in Section~\ref{sec:AshapeSFH}, these trends indicate that strongly disturbed galaxy pairs at intermediate projected separations had stronger star formation in the recent past relative to their current activity. 

The right panel shows the corresponding behavior of $\Delta \dn_{\ha}$. Galaxies with low $\ashape$ show little variation with projected separation and remain close to the control sample. In contrast, galaxies with high $\ashape$ show a clear decrease in $\Delta \dn_{\ha}$ at intermediate separations ($\sim100$--$150\kpc$), indicating younger stellar populations than isolated galaxies with similar current star formation activity. This suggests that these systems formed a larger fraction of their stellar populations through enhanced star formation over the past $\sim1{-}2$ Gyr

Taken together, these results indicate that strongly disturbed galaxies at intermediate projected separations experienced stronger star formation in the recent past than isolated galaxies with similar current star formation activity. The differences between the current and longer-timescale diagnostics provide further evidence that their star formation activity has changed significantly over the past $\sim1{-}2$ Gyr.

\subsection{Physical Interpretation}
\label{sec:PhysicalInterpretation}

Numerical simulations show that galaxy interactions are strongly time-dependent. Galaxy pairs typically undergo multiple orbital passages before final coalescence, including close encounters, subsequent motion toward larger separations, and later reapproaching phases \citep[e.g.,][]{Mihos1996, Springel2005, Sparre2022}. Near close passage, tidal perturbations can produce asymmetric structures such as tidal tails, bridges, and distorted outer light distributions \citep{Toomre1972, Lotz2008}. Tidal torques can also drive gas inflows toward galaxy centers and enhance star formation \citep{Cox2008, Torrey2012}. As the galaxies subsequently move apart, tidal features may remain visible and gradually weaken because of dynamical relaxation \citep{Lotz2008}, while the interaction-induced star formation may decline as gas inflows weaken and the available cold gas is consumed \citep{Montuori2010}. Motivated by these results, we consider a possible evolutionary interpretation of our observations.

In this picture, strongly disturbed pairs at small projected separations may represent systems near a close passage. Their strong enhancements in $\logsfr$ and $\haew$, together with relatively weak $\hdew$ enhancement, suggest that their current star formation has risen recently. Strongly disturbed galaxies at intermediate projected separations may instead include systems moving away after a close passage. Their current star formation is weaker, while enhanced $\hdew$ and reduced $\dn$ indicate stronger star formation in the recent past. A possible explanation is therefore that SFR rises rapidly around close passage and then declines as the galaxies move apart, while tidal features and longer-timescale stellar-population signatures remain visible.

The low-$\ashape$ population is likely more heterogeneous and may include both weakly interacting systems and galaxies whose tidal features have faded after an earlier encounter. More generally, projected separation and morphological disturbance cannot uniquely determine whether a pair is approaching or moving away from close passage because of projection effects and the non-monotonic evolution of pair separation \citep{Perez2006, Contreras-Santos2022, Patton2024}. The evolutionary picture described above is therefore only one possible interpretation of the observations. In Section~\ref{sec:dis}, we test whether this picture arises in the time-resolved orbital, morphological, and star formation histories of TNG100 galaxy pairs.

\section{Interpreting the Observed Trends with Cosmological Simulations} \label{sec:dis}

To test the evolutionary picture proposed in Section~\ref{sec:PhysicalInterpretation}, we compare the observed trends with galaxy pairs from cosmological hydrodynamical simulations. The time-resolved orbital, morphological, and star formation histories allow us to examine whether the observed combinations of projected separation, morphological disturbance, and recent star formation indicators arise naturally during galaxy interactions.

\subsection{IllustrisTNG Simulation}

The IllustrisTNG project \citep{Pillepich2018b, Nelson2019} is a suite of cosmological magneto-hydrodynamical simulations performed with the moving-mesh code \textsc{AREPO} \citep{Springel2010}. The simulations include a comprehensive galaxy formation model incorporating gas cooling, star formation, chemical enrichment, and stellar and AGN feedback \citep{Weinberger2017, Pillepich2018a}, enabling them to reproduce a broad range of observed galaxy properties.

The IllustrisTNG suite consists of three main simulation volumes with side lengths of 300, 100, and 50 cMpc. In this work, we use the TNG100 simulation, which provides a suitable balance between volume and spatial resolution for studying galaxy interactions. TNG100 has a dark matter particle mass of $7.5\times10^6\ M_\odot$, a baryonic mass resolution of $1.4\times10^6\ M_\odot$, and a stellar gravitational softening length of $\sim0.74$ kpc at $z=0$.

Galaxies in IllustrisTNG are identified as gravitationally bound subhalos using the \textsc{SUBFIND} algorithm \citep{Springel2001}, while their evolutionary histories are reconstructed with the \textsc{SUBLINK} merger trees \citep{Rodriguez-Gomez2015}. This framework provides time-resolved information on galaxy orbits, star formation histories, and stellar mass assembly, allowing the interaction history of galaxy pairs to be directly tracked across cosmic time. These properties make TNG100 particularly well suited for this work, as it enables us to investigate how galaxy separation, morphological disturbance, and star formation activity co-evolve throughout the merger process.

\subsection{Selection of Galaxy Pairs in TNG100}

\begin{figure*}
    \centering
    \includegraphics[width=\linewidth]{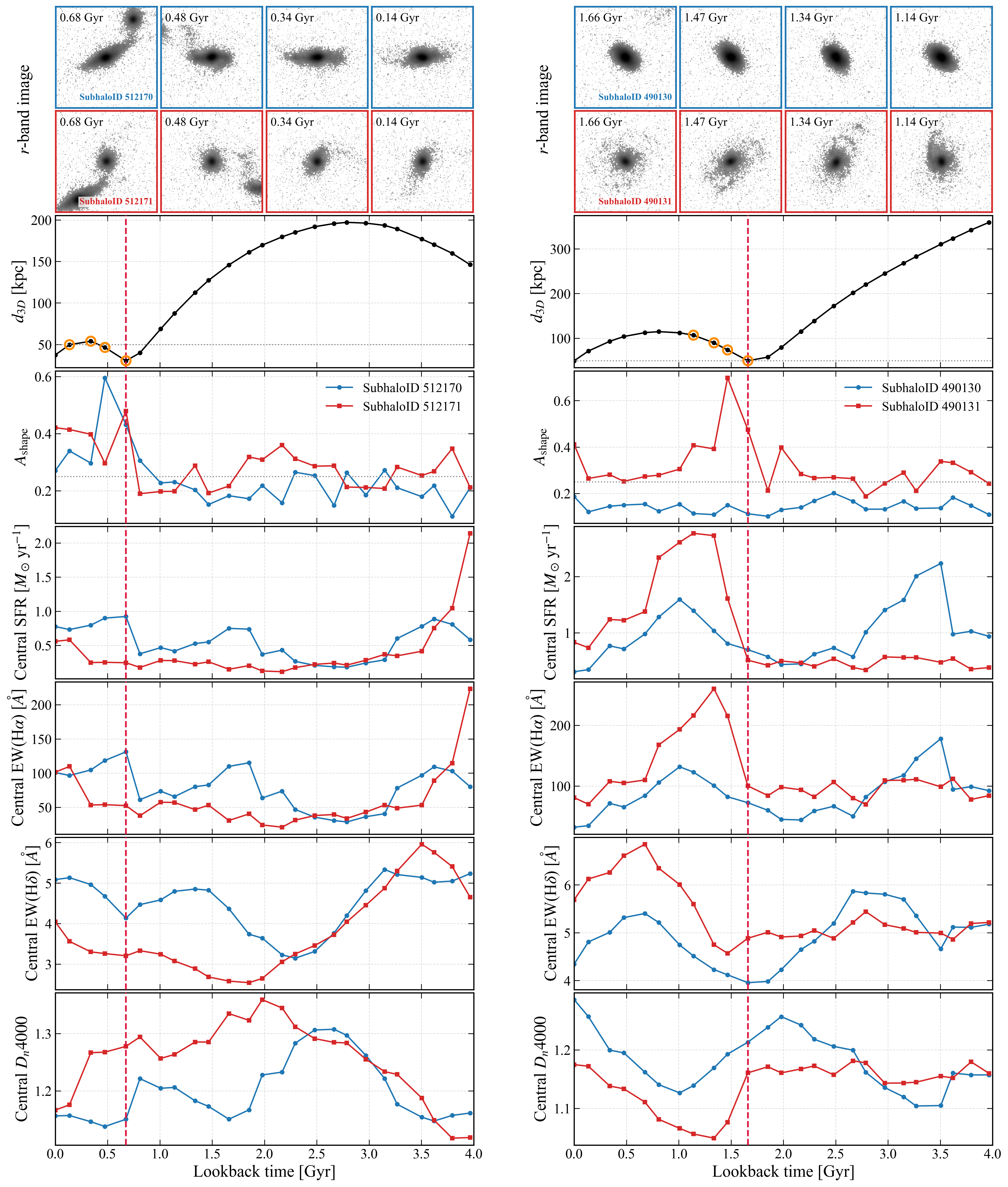}
    \caption{
    Evolution of two example TNG100 pairs over the past $\sim4$ Gyr. The left and right columns correspond to the two pairs, respectively. The top row shows mock $r$-band images of the two pair members at the identified close-passage snapshot and the following three snapshots. The remaining rows show, from top to bottom, the three-dimensional pair separation, $\ashape$, SFR, $\haew$, $\hdew$, and $\dn$ as functions of lookback time. The SFR and spectral indices are measured within one stellar half-mass radius of each pair member. The vertical red dashed line marks the snapshot at which $d_{\rm 3D}$ reaches a local minimum below $50\ {\rm kpc}$, providing an approximate estimate of the recent close-passage epoch.
    }
    \label{fig:MockPairTback}
\end{figure*}

To construct a TNG100-1 pair sample comparable to the observed isolated star-forming spiral--spiral major-merger sample described in Section~\ref{sec:pair_sample}, we select pairs of subhalos at $z=0$ with stellar masses of $9.5\leq\log(M_\star/M_\odot)\leq11.0$. Both members are required to be late-type systems, adopting $P_{\rm Late}>0.5$ from the TNG100 deep-learning morphology catalog \citep{Huertas-Company2019, Varma2022}. We further require a stellar-mass ratio of $1\leq M_{\star,1}/M_{\star,2}\leq3$, consistent with the major-merger definition adopted for the observational sample, where member 1 denotes the more massive subhalo. Taking advantage of the full three-dimensional information available in the simulation, we select subhalo pairs with $d_{\rm 3D}<200\ {\rm kpc}$.

We further select relatively isolated systems to reduce the influence of a more complex group environment. Specifically, for each member, we require that, excluding its paired companion, no additional subhalo within a three-dimensional radius of $200\ {\rm kpc}$ has a stellar mass exceeding $10\%$ of that member's stellar mass. Applying these stellar-mass, morphology, mass-ratio, separation, and isolation criteria yields $74$ isolated late-type major-pair candidates at $z=0$.

We trace the recent evolution of both members of each $z=0$ subhalo pair along their main progenitor branches using the \textsc{SubLink} merger trees. The histories are reconstructed over the preceding $4\ {\rm Gyr}$ to characterize their recent evolution. We retain only systems in which both members remain star-forming throughout this interval, with each member satisfying $\log(\mathrm{sSFR}/\mathrm{yr}^{-1})>-11$ at every snapshot along its main progenitor branch. This requirement keeps the simulated sample comparable to the star-forming spiral--spiral pairs in the observational sample and reduces the influence of other mechanisms that can suppress star formation, such as hot circumgalactic gas associated with quenched galaxies \citep{Feng2024, Shi2026}.

We then identify systems that have experienced a recent close passage. Specifically, we require the three-dimensional pair separation to reach a local minimum below $50\ {\rm kpc}$ within the past $2\ {\rm Gyr}$, with the separation increasing again afterward. Because the orbital histories are sampled only at discrete simulation snapshots, the snapshot corresponding to this local minimum provides only an approximate estimate of the true pericentric passage \citep{Patton2024}. Hereafter, we refer to this snapshot as the \textit{close-passage epoch}. We further require at least three snapshots after the close-passage epoch to ensure sufficient sampling of the subsequent evolution in star formation and morphology. These criteria yield a final sample of $12$ recently interacting, star-forming galaxy pairs. Hereafter, we refer to each selected two-subhalo system as a \textit{TNG100 pair}, and to the two subhalos and their main progenitors as the \textit{pair members}.

Figure~\ref{fig:MockPairTback} shows the recent evolution of two example pairs from the final TNG100 pair sample. The second row presents their three-dimensional separation as a function of lookback time, with the close-passage epoch marked by the vertical red dashed line. In both systems, the pair members approach each other and subsequently move apart. As required by our selection criteria, more than three snapshots are available after the close-passage epoch, allowing the post-passage separation phase to be adequately sampled.

\subsection{Evolution of Shape Asymmetry during Galaxy Interactions}

For each pair and snapshot, we use the stellar particles associated with the two subhalos to construct mock $r$-band images. We adopt a fixed $x$--$y$ projection, with the $z$-axis treated as the line of sight, and generate separate images centered on each pair member while including the stellar particles of both subhalos in the field of view. We use $d_{\rm proj}$ to denote the separation between the two pair members projected onto this image plane. Each image covers a fixed physical scale of approximately $50\ {\rm kpc}$.

Because particle-level photometry is not uniformly available across all snapshots, we convert stellar mass to $r$-band luminosity using a fixed mass-to-light ratio of $M_\star/L_r=1.5\ M_\odot/L_{\odot,r}$. This value is representative of the relatively blue stellar populations of star-forming late-type galaxies \citep[e.g.,][]{Tojeiro2013, Bell2003}. This simplified treatment does not account for spatial variations in stellar populations, metallicity, or dust attenuation. We also tested modest variations in the adopted mass-to-light ratio and found no significant change in the resulting $\ashape$ trends.

To make the observational effects in the mock images similar to those in the DESI Legacy Surveys, all systems are placed at a common mock-observation redshift of $z_{\rm mock}=0.05$. The mock images are generated with a pixel scale of $0.262''$ pixel$^{-1}$, convolved with a Gaussian point-spread function of ${\rm FWHM}=1.5''$, and noise is added to reproduce a point-source $5\sigma$ depth of $r=23.9$ mag.

We measure $A_{\rm shape}$ from the mock images using the same source-detection, segmentation, masking, and $180^\circ$-rotation procedure applied to the observational images, as described in Section~\ref{sec:ashape}. We also adopt the same surface-brightness thresholds and segmentation parameters so that the mock and observed images are analyzed with the same measurement procedure.

The first row of Figure~\ref{fig:MockPairTback} shows mock $r$-band images of the two members in each pair at the identified close-passage snapshot and the following three snapshots. These four epochs are marked by circles on the corresponding $d_{\rm 3D}$ evolution curves. The third row shows the measured $\ashape$ values and their evolution with lookback time. Among the four pair members, three develop clear tidal features near the close-passage epoch, which remain visible to some extent during the subsequent separation phase. Their $\ashape$ values increase accordingly, indicating that morphological disturbance is strongly enhanced around close passage and can persist afterward, consistent with previous numerical studies of interacting galaxies \citep{McElroy2022}.

One pair member, however, remains relatively regular in morphology during the close passage and shows little variation in $\ashape$. This demonstrates that a close encounter does not necessarily produce strong visible tidal features in every individual galaxy, and that the morphological response can depend on other properties of the interaction \citep{Lotz2008, Lotz2010b}. In several other pairs not shown here, irregular morphologies are also present at relatively large pair separations, which may arise from minor mergers or other internal processes. Distinguishing these additional sources of morphological disturbance is beyond the scope of this work, which focuses on the evolution of galaxy properties around the identified close-passage events.

These examples illustrate that tidal features and $\ashape$ provide useful tracers of recent interactions, but the relation between morphological disturbance and orbital stage has substantial scatter for individual systems. They are therefore more reliable as statistical indicators of interaction history than as precise tracers of the merger stage for individual galaxies.

\subsection{Evolution of Spectral Indices during Galaxy Interactions}
\label{sec:fsps}

For each pair member, we trace the recent star formation history along its main progenitor branch over the preceding $4\ {\rm Gyr}$. To compare with the SDSS fiber measurements, we use the SFR within the three-dimensional stellar half-mass radius, \texttt{SubhaloSFRinHalfRad}, as the central SFR. This aperture is not identical to the fixed-angular-size SDSS fiber, but it samples the central region of each simulated galaxy and therefore provides a closer comparison with the observations.

Because the recent $4\ {\rm Gyr}$ are most relevant for tracing the evolution associated with the identified close passages, we reconstruct the star formation history in detail over this interval and adopt a simplified prescription for earlier times. We assume a constant SFR from a cosmic age of $0.5\ {\rm Gyr}$ to the beginning of the reconstructed TNG history, normalized to the median of the five earliest SFR measurements in the recent history. This earlier component mainly provides the underlying old stellar population required for the spectral synthesis. Since our analysis focuses on the relative evolution of the spectral diagnostics around close passage rather than their absolute values, the precise form of the earlier star formation history has little effect on the trends of interest. We have also tested alternative assumptions for the earlier star formation history and find that the resulting evolutionary trends remain unchanged.

We generate synthetic spectra from the resulting star formation histories using \texttt{FSPS} (Flexible Stellar Population Synthesis; \citealt{Conroy2009, Conroy2010}) through \texttt{python-fsps} \citep{ForemanMackey2014}. We adopt a Chabrier initial mass function and the default FSPS stellar-evolution prescriptions and spectral libraries. The tabulated star formation histories are interpolated between the discrete TNG snapshots, and spectra are generated at the cosmic age corresponding to each snapshot. For each snapshot, we generate both a stellar-only spectrum and a spectrum including nebular emission. The two spectra are then used together to measure the emission-line and stellar-continuum diagnostics, including H$\alpha$, H$\delta_A$, and $D_n4000$. Dust attenuation is not included.

The lower four rows of Figure~\ref{fig:MockPairTback} show the evolution of the central SFR, $\haew$, $\hdew$, and $\dn$ over the past $\sim4$ Gyr. Among the four pair members, three show a clear enhancement in central SFR following the identified close passage, accompanied by a similar increase in $\haew$. The peak in SFR does not always coincide exactly with the minimum in $d_{\rm 3D}$, and in some cases occurs shortly afterward. The evolution of $\hdew$ is noticeably delayed relative to the SFR and $\haew$. Its enhancement becomes strongest after the star formation peak, when the pairs have already moved to larger separations and in some cases reached the apocentric phase of their orbit. In contrast, $\dn$ decreases rapidly during the interaction-induced star formation enhancement because of the increasing contribution from young stellar populations, and subsequently evolves more gradually on longer timescales.

These examples demonstrate that the evolutionary picture proposed in Section~\ref{sec:PhysicalInterpretation} can produce the combinations of current and past star formation indicators seen in the observations. In particular, a pair can show strong $\hdew$ enhancement and reduced $\dn$ at intermediate or large separations after its current star formation enhancement has begun to decline.

However, the response is not identical for every galaxy. One of the four pair members shown in Figure~\ref{fig:MockPairTback} shows no immediate enhancement in central SFR. This indicates substantial galaxy-to-galaxy variation in the response to interactions. We therefore extend the analysis to all selected pairs and snapshots to test whether the trends seen in these examples are recovered statistically.

\subsection{Spectral Indices as a Function of Projected Pair Separation}

\begin{figure*}
    \centering
    \includegraphics[width=\linewidth]{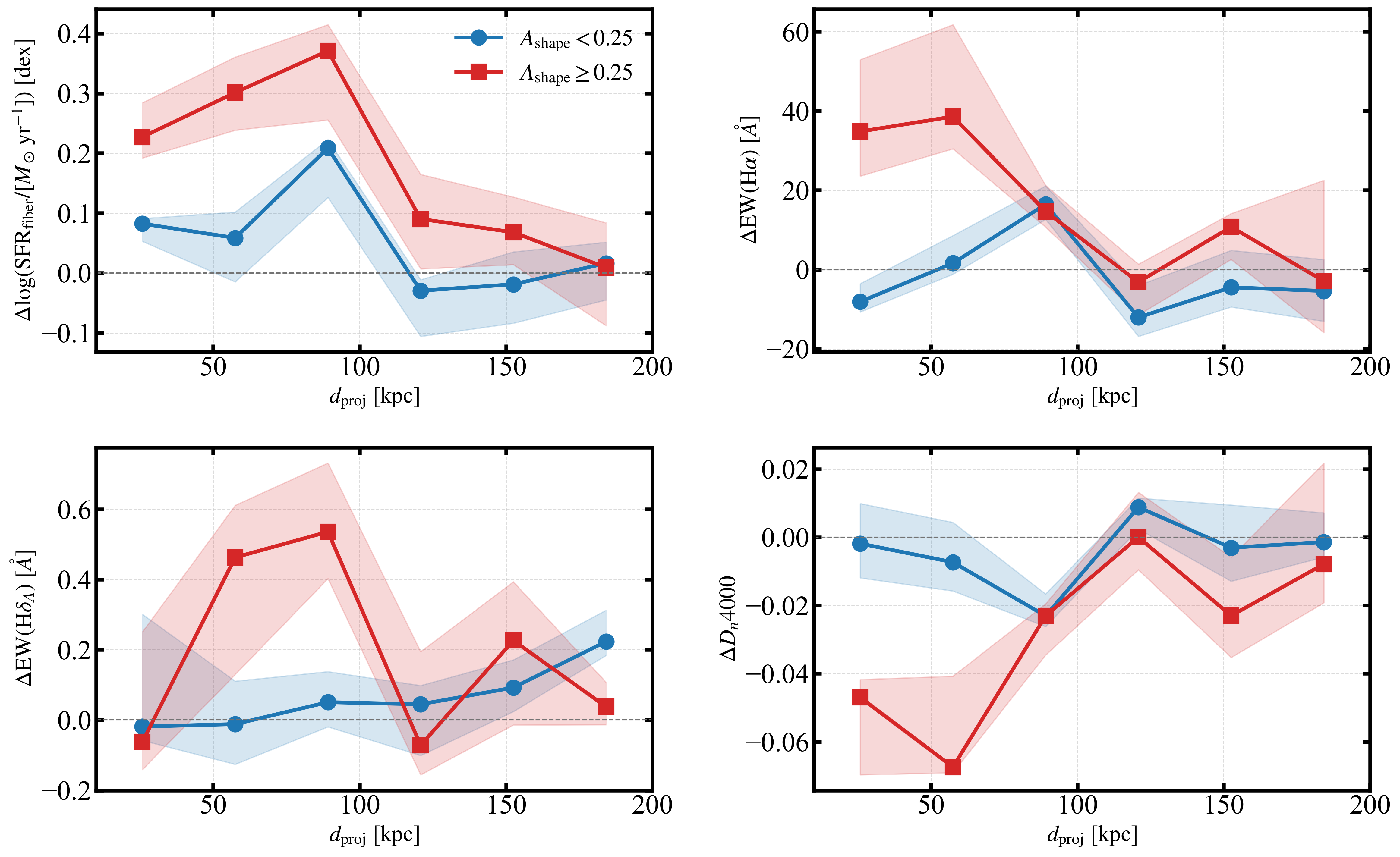}
    \caption{Same as Figure~\ref{fig:AshapeSFR}, but for the $406$ mock galaxies obtained by tracing both members of the $12$ selected TNG100 pairs over the past $4\ {\rm Gyr}$ and retaining snapshots with $10<d_{\rm proj}<200\ {\rm kpc}$. For each pair member, the offset in each star formation diagnostic is defined as the value at the current snapshot minus the median value over snapshots before close passage with $d_{\rm 3D}>150\kpc$. Red squares and blue circles represent mock galaxies with $\ashape\geq0.25$ and $\ashape<0.25$, respectively.}
    \label{fig:MockPairDp}
\end{figure*}

To make a statistical comparison with the observations, we trace both members of the $12$ selected $z=0$ TNG100 pairs through all available snapshots over the past $4\ {\rm Gyr}$. Each pair member at each snapshot is treated as one mock galaxy. The same pair member can therefore contribute multiple mock galaxies at different evolutionary times. For each mock galaxy, we record the projected pair separation, $\ashape$, central SFR, and spectral indices, and restrict the analysis to $10<d_{\rm proj}<200\ {\rm kpc}$, matching the separation range adopted for the observational sample. This yields $406$ mock galaxies for the statistical analysis; they represent $406$ data points rather than $406$ distinct subhalos. As in the observational analysis, we focus on the offsets in star formation and spectral properties relative to the pre-passage large-separation stage. For each pair member, the reference value is taken as the median measured before close passage at $d_{\rm 3D}>150\ {\rm kpc}$, and the corresponding offset is calculated for each snapshot.

The $406$ mock galaxies are then divided into low- and high-asymmetry populations using $\ashape=0.25$. Figure~\ref{fig:MockPairDp} shows the median offsets as a function of projected separation for the two populations. The small number of mock pairs leads to relatively large statistical uncertainties, while the simplified assumptions used to generate the mock data can produce differences from the observations in the detailed trends. Nevertheless, the overall behavior is consistent with that observed in the real galaxy pairs. Mock galaxies with high $\ashape$ show stronger enhancements in both $\Delta \logsfr$ and $\Delta \haew$ than those with low $\ashape$, and both quantities increase toward smaller projected separations. High-$\ashape$ mock galaxies also show lower $\Delta \dn$ than low-$\ashape$ mock galaxies, with $\Delta \dn$ decreasing further toward smaller $d_{\rm proj}$. For mock galaxies with low $\ashape$, $\Delta \hdew$ shows little dependence on projected separation. In contrast, high-$\ashape$ mock galaxies show a strong $\Delta \hdew$ enhancement at $d_{\rm proj}\sim50$--$100\kpc$, where it is substantially higher than in the low-$\ashape$ population.

Following the approach illustrated in Figure~\ref{fig:MockPairTback}, we examine the evolutionary histories of all 12 TNG100 pairs. In most pairs, the central SFR begins to rise rapidly around close passage and then gradually declines as the galaxies move apart. About half of the pair members also develop clear tidal features after close passage. These evolutionary trends give rise to the statistical correlations among $\ashape$, recent star formation history, and projected pair separation shown in Figure~\ref{fig:MockPairDp}, further supporting the evolutionary picture proposed in Section~\ref{sec:PhysicalInterpretation}.

At the same time, the TNG100 pairs show that not every interacting galaxy follows the same evolutionary trend. The relations among morphological disturbance, projected separation, and recent star formation history should therefore be interpreted statistically. The inferred SFR evolution represents an average trend for star-forming spiral--spiral major mergers rather than a unique evolutionary path for every system.

\section{Summary}\label{sec:sum}

In this work, we jointly investigate morphological disturbance, projected pair separation, and recent star formation history in galaxy pairs to observationally constrain how star formation evolves along the merger sequence. Our analysis is based on a sample of $5090$ star-forming galaxies in spiral--spiral major-merger pairs selected from the SDSS main galaxy sample, spanning a redshift range of $0.02<z<0.12$ and a stellar mass range of $9.5<\log(M_*/M_\odot)<11.0$.

Using deep imaging from the DESI Legacy Surveys, we measure the shape asymmetry parameter ($\ashape$) for the pair galaxies. Comparison with the Galaxy Zoo DESI morphology classifications shows that galaxies with prominent tidal disturbances generally have higher $\ashape$. We also find that the median $\ashape$ increases toward smaller projected pair separations, while showing little dependence on stellar mass or redshift over the ranges considered here. These results support the use of $\ashape$ as a quantitative indicator of tidal disturbance in interacting galaxies.

We then compare $\logsfr$ and $\haew$, which reflect current star formation activity, with $\hdew$ and $\dn$, which are sensitive to stellar populations formed over longer timescales. Their dependence on both $\ashape$ and projected pair separation allows us to constrain changes in recent star formation history. Our main observational results are summarized as follows:

\begin{enumerate}

\item Galaxies with high $\ashape$ show stronger enhancements in $\logsfr$ and $\haew$ than those with low $\ashape$, with the enhancements increasing toward smaller projected separations. Their $\dn$ values are also generally lower, indicating younger stellar populations. In contrast, the strongest $\hdew$ enhancement occurs at intermediate projected separations of $\sim50$--$100\kpc$, rather than at the smallest separations. This indicates that strongly disturbed galaxies at intermediate separations experienced stronger star formation over the past $\sim0.1$--$1$ Gyr, whereas those at the smallest separations are undergoing the strongest current star formation. This conclusion is further supported by comparisons with isolated galaxies matched in current star formation activity.

\item Galaxies with low $\ashape$ also show enhanced $\logsfr$ and $\haew$, together with reduced $\dn$, toward smaller projected separations, although these trends are weaker than those in high-$\ashape$ galaxies. In contrast, their $\hdew$ remains close to that of isolated galaxies over most projected separations and shows only a mild enhancement at the smallest separations. These results indicate that galaxies without strong morphological disturbances can still experience enhanced current star formation during close interactions, while the mild $\hdew$ enhancement at the smallest separations suggests somewhat stronger star formation over the past $\sim0.1$--$1$ Gyr.

\item The Galaxy Zoo DESI classifications yield results consistent with those based on $\ashape$. Galaxies classified as \textit{disturb} show stronger current star formation at small projected separations, lower $\dn$, and significant $\hdew$ enhancement mainly at relatively large projected separations, further supporting the connection between morphological disturbance and recent star formation history.

\end{enumerate}

To test the evolutionary picture inferred from the observations, we trace 12 recently interacting galaxy pairs selected from TNG100. Their evolutionary histories show that the central SFR generally rises rapidly around close passage and declines as the galaxies subsequently separate, while tidal disturbances and longer-timescale stellar-population signatures persist. Despite the small pair sample and the simplified mock-observation procedure, the resulting mock galaxies broadly reproduce the observed relations among projected separation, $\ashape$, and the star formation diagnostics. These results support the interpretation that the strong $\hdew$ enhancement in disturbed galaxies at intermediate separations reflects the delayed spectral response to stronger star formation around an earlier close passage.

\section*{Acknowlegements}
We thank the anonymous referee for the constructive comments and suggestions that improved this paper. This work is supported by the National Natural Science Foundation of China (No. 12103017), the Natural Science Foundation of Hebei Province (No. A2025205037). SZG, SHC, YTD and YJN acknowledge support from the Undergraduate Innovation and Entrepreneurship Training Program of Hebei Normal University (No. 202310094009 and X202610094074). CLS acknowledges support from the Graduate Innovation Fund of Hebei Normal University (No. XCXZZBS202541).

This work made use of data from the Sloan Digital Sky Survey (SDSS). Funding for SDSS has been provided by the Alfred P. Sloan Foundation, the U.S. Department of Energy Office of Science, and the Participating Institutions. The SDSS website is \url{https://www.sdss.org}.

This work made use of data from LAMOST (Large Sky Area Multi-Object Fiber Spectroscopic Telescope, also known as the Guoshoujing Telescope) (\url{https://cstr.cn/31118.02.LAMOST}). LAMOST is a Chinese national mega-science facility operated by the National Astronomical Observatories, Chinese Academy of Sciences.

This work also made use of data products from the Legacy Surveys. The Legacy Surveys consist of three individual and complementary projects: the Dark Energy Camera Legacy Survey (DECaLS; Proposal ID \#2014B-0404; PIs: David Schlegel and Arjun Dey), the Beijing-Arizona Sky Survey (BASS; NOAO Prop. ID \#2015A-0801; PIs: Zhou Xu and Xiaohui Fan), and the Mayall z-band Legacy Survey (MzLS; Prop. ID \#2016A-0453; PI: Arjun Dey). DECaLS, BASS, and MzLS together include data obtained, respectively, at the Blanco telescope, Cerro Tololo Inter-American Observatory, NSF's NOIRLab; the Bok telescope, Steward Observatory, University of Arizona; and the Mayall telescope, Kitt Peak National Observatory, NOIRLab. Pipeline processing and analyses of the data were supported by NOIRLab and the Lawrence Berkeley National Laboratory (LBNL). The Legacy Surveys project is honored to be permitted to conduct astronomical research on Iolkam Du'ag (Kitt Peak), a mountain with particular significance to the Tohono O'odham Nation.

\bibliography{ref}
\bibliographystyle{aasjournal}

\end{document}